\documentclass[a4paper]{jpconf}
\usepackage{amsmath,amssymb,epsfig,amsfonts,multirow,tabu,iopams,pdfpages,color,enumitem,titlesec}

\usepackage{titling}
\usepackage{braket}
\usepackage{blkarray}
\usepackage{lipsum}
\usepackage[section]{placeins}
\usepackage{bbold}
\usepackage[normalem]{ulem}
\usepackage[font=footnotesize,labelfont=bf, justification=centering]{caption}
\usepackage{graphicx,subfigure}
\usepackage{hyperref}
\usepackage[all]{hypcap}
\makeatletter

\newcommand{\Romannum}[1]{\expandafter\@slowromancap\romannumeral #1@}

\numberwithin{equation}{section}
\newcommand{\bes}{\begin{split}}
\newcommand{\ees}{\end{split}}
\def\ie{{\it i.e.~}}
\def\eg{{\it e.g.~}}

\newcommand{\bv}[1]{\mathbf{b}_{#1}}

\def\notE6{{$SO(10)\times U(1)_{\zeta}\not\subset E_6$}}
\def\E6{{$SO(10)\times U(1)_{\zeta}\subset E_6$}}

\def\highSO10{{$SU(3)_{C}\times SU(2)_{\!L} \times SU(2)_{\!R} \times U(1)_{C}$}}

\let\originalleft\left
\let\originalright\right
\renewcommand{\left}{\mathopen{}\mathclose\bgroup\originalleft}
\renewcommand{\right}{\aftergroup\egroup\originalright}

\begin{document}
\title{Light $U(1)$s in heterotic string models}

\author{V M Mehta}

\address{Institute for Theoretical Physics,\\
  University of Heidelberg,\\
  Philosophenweg 16,\\
  69120 Heidelberg,\\
Germany}
\ead{viraf.mehta@thphys.uni-heidelberg.de}

\begin{abstract}
We present a novel string-derived $U(1)$ combination that satisfies necessary properties to survive to low scales.  We discuss previous attempts at acquiring such an abelian gauge symmetry from two different string embeddings and the pitfalls associated with them.  Finally, we give an example of how a satisfactory model may be constructed within our framework.
\end{abstract}

\section{Introduction}
Evidence for additional gauge symmetries has, so far, been absent in experimental searches.  However, the theoretical motivations of going beyond the standard model gauge group are abundant (see \eg \cite{Langacker:2010zza} and references therein).  An important consideration, when going beyond the standard model, is the question of proton stability.  As soon as one begins supersymmetric model building, dimension-4 and -5 proton decay mediating operators are induced.  

Various attempted resolutions have been proposed in the literature, both from bottom-up and top-down perspectives: gauged $B$- and $L$-numbers\cite{FileviezPerez:2010gw}; gauged $B-L$\cite{Faraggi:1990ita,Braun:2005ux} and; a variety of discrete symmetries\cite{Kobayashi:2006wq,Mohapatra:2007vd}.  However, some of these additional symmetries have consequences that make realistic model building problematic\cite{Faraggi:1990it}.  

In this work, we attempt to acquire an additional abelian gauge symmetry from string models.  As we begin at the string scale, it is useful to construct a list of properties such a gauge symmetry should satisfy in order to be a viable low-scale gauge symmetry.  This list is not, by any means, exhaustive, however provides some guidance in our approach\cite{Faraggi:2011xu,Faraggi:2013nia}:

\begin{itemize}
  \item To sufficiently suppress or forbid proton decay mediating operators upto dimension-6;
  \item To allow for some mechanism for $m_\nu\sim 1$eV;
  \item To have the electroweak Yukawa couplings invariant;
  \item To be family universal\footnote{this is not a necessary requirement, however we choose to impose it to avoid proton decay mediation via intrafamily operators};
  \item To be free of gauge and gravitational anomalies;
  \item To satisfy low-scale gauge coupling data.
\end{itemize}

During this talk, we will introduce the framework with which we build our model and then briefly discuss examples of $U(1)$s that were of interest to us in the past.  We then present a novel string-derived model that has the necessary properties to facilitate the above requirements.

\section{Heterotic model building}
The free fermionic formulation \cite{Kawai:1986ah,Antoniadis:1986rn,Antoniadis:1987wp} allows for the construction of string models directly in four dimensions and cancels the conformal anomaly by introducing freely propagating, fermionic, worldsheet degrees of freedom, detailed in Table \ref{table:wsfermions} for the heterotic string.  
\begin{table}[!h]
\noindent 
{\small
\begin{center}
{\tabulinesep=1.5mm
\begin{tabu}{|c|cc|p{9.5cm}|}
\hline\hline
\textbf{Sector}&\multicolumn{2}{c|}{\textbf{Label}} &\textbf{Description} \\
\hline
\multirow{5}{*}{SUSY}& &$\psi^\mu$& Majorana--Weyl superpartners of the bosonic coordinates \\
\cline{2-3}
& &$\chi^{i}$&  Majorana--Weyl superpartners to the six compactified dimensions \\
& &$y^{i},w^{i}$& Majorana--Weyl fermions that correspond to the bosons describing the six compactified dimensions in the bosonic formulation \\
\hline\hline
\multirow{5}{*}{Non-SUSY}& &$\overline{y}^{i},\overline{w}^{i}$& Majorana--Weyl fermions that correspond to the bosons describing the six compactified dimensions  in the orbifold formulation \\
&\multirow{2}{*}{$\overline{\lambda}^I\left\{\vphantom{\begin{array}{c}1\\1\\1\\1\end{array}}\right.$} &$\overline{\psi}^{1,\dots,5},\overline{\eta}^{1,2,3}$& Complex fermions that describe the visible gauge sector, corresponding to eight coordinates of the internal $T^{16}$  \\
& &$\overline{\phi}^{1,\dots,8}$& Complex fermions that describe the hidden gauge sector, corresponding to the remaining eight coordinates parametrising the internal $T^{16}$ \\
\hline\hline
\end{tabu}}\end{center}
}
\caption{\label{table:wsfermions}
This table gives the fermionic states that freely propagate on the string worldsheet with 
$\mu=1,2$, $i=1,\dots, 6$ and $I=1,\dots, 16$, as four dimensional light-cone, six real internal and sixteen complex indices, respectively. }
\end{table}

\noindent Unlike worldsheet bosonic degrees of freedom, worldsheet fermions, generically, pick up a phase on parallel transport around the non-contractible loops of the toroidal worldsheet,
\begin{align}
  f\rightarrow -e^{i\pi\alpha\left( f \right)}f,
\end{align}
where we restrict ourselves to $\alpha\left( f \right)\in\left(-1,+1\right]$.  64-component basis vectors, $\mathbf{b}_i$, describing the phases of the fermions, span a finite additive set, $\Xi$, where 
\begin{align}\bes
  \Xi&=\mathrm{span}\left\{ \mathbf{b}_1,\dots,\mathbf{b}_k \right\}\\
  &\simeq \mathbb{Z}_{N_1}\oplus\cdots\oplus\mathbb{Z}_{N_k}.
\label{additivegroup}
\ees
\end{align}
The elements of this additive set make up the various sectors of our string model, $\alpha=\sum m_i\mathbf{b}_i$, such that
\begin{align}
  m_i\mathbf{b}_i=0\ \mathrm{mod}\ 2 \Leftrightarrow m_i=0\ \mathrm{mod}\  N_i, \forall i=1,\dots,k,
\end{align}
and we can write the partition function as 
\begin{align}
  \mathcal{Z}=\sum_{\alpha,\beta\in\Xi}c\left(^\alpha_\beta \right)Z\left[^\alpha_\beta \right].
\end{align}
Thus, free fermionic models can be fully described using the two sets of input parameters: the set of basis vectors, $\mathcal{B}=\left\{ \mathbf{b}_i \right\}$ and the corresponding GSO projections, $c(^\alpha_\beta)$.  Modular invariance is then guaranteed by satisfying the \textit{ABK rules} \cite{Antoniadis:1986rn}.

\subsection{$U(1)$s in free fermion models}
The gauge group generated in free fermion models is rank-22 with the Cartan subalgebra generated by the right-moving currents $\overline{f}^\ast\overline{f}$.  The corresponding $U(1)$ charges for each complex fermion, $f$, are given by
\begin{align}
  Q(f)=\frac 12 \alpha\left( f \right)+F\left( f \right),
\end{align}
where $F(f)$ is the \textit{fermion number} of $f$.  

For massless sectors consisting of periodic complex fermions only, the vacuum is degenerate, $\ket{\pm}$.  These represent the Clifford algebra generated by the zero modes, $f_0$ and $f^\ast_0$, which have $F(f)=0,1$, respectively, and thus, have $Q\left( \ket{\pm} \right)=\pm\frac 12$.

The $U(1)$ symmetries of interest in this talk are those generated by $\overline{\eta}^{i\ast}\overline{\eta}^i$, $i=1,2,3$, which correspond to the Cartan generators of our choice of observable gauge group, along with $\overline{\psi}^{j\ast}\overline{\psi}^{j}$, $j=1,\dots,5$, generating the $SO(10)$ GUT.  

Early heterotic string models, constructed using the free fermionic formulation, were NAHE-based; that is, additional basis vectors were added to the canonical set $\mathcal{B}_{\mathrm{\tiny{NAHE}}}=\left\{ \mathbb{1},\mathbf{S},\mathbf{b}_1,\mathbf{b}_2,\mathbf{b}_3 \right\}$ to build semi-realistic string models, with
\begin{align}\begin{split}
  \mathbb{1}&=\left\{ \mathrm{All} \right\},\\
  \mathbf{S}&=\left\{ \psi^\mu,\chi^{1,\dots,6} \right\},\\
 \mathbf{b}_1&=\left\{\psi^\mu,\chi^{12}, y^{3,\dots,6}\overline{y}^{3,\dots,6},\overline{\psi}^{1,\dots,5},\overline{\eta}^{1}\right\},\\
\mathbf{b}_2&=\left\{\psi^\mu,\chi^{34}, y^{1,2},w^{5,6}\overline{y}^{1,2},\overline{w}^{5,6},\overline{\psi}^{1,\dots,5},\overline{\eta}^{2}\right\},\\
\mathbf{b}_3&=\left\{\psi^\mu,\chi^{56}, w^{1,\dots,4}\overline{w}^{1,\dots,4},\overline{\psi}^{1,\dots,5},\overline{\eta}^{3}\right\},
\end{split}\end{align}  
 where fermions in $\left\{ \dots \right\}$ are periodic.  This, with the correct choice of GGSO phases, results in an $\mathcal{N}=1$ supersymmetric $SO(10)\times SO(6)^3\times E_8$ gauge group in four dimensions, with 48 generations of the spinor-$\mathbf{16}$ representation of $SO(10)$.  Upon addition of further basis vectors, the number of generations is reduced and the $SO(6)^3$ is broken to $U(1)$ factors, and the remaining SUSY maybe broken.  However, in the following, we choose to preserve the $\mathcal{N}=1$ SUSY.  The $SO(10)$ GUT is also broken to one of its subgroups.  

 \section{Symmetry breaking patterns and embeddings}

 There are two classes of $SO(10)$ breakings, as previously constructed in the literature using the free fermionic formulation of the heterotic string:
 \begin{enumerate}
   \item Flipped $SU(5)$\cite{Antoniadis:1988tt}, Pati-Salam\cite{Antoniadis:1990hb} and Standard-like models\cite{Faraggi:1989ka,Faraggi:1991jr};
   \item Left-right symmetric\cite{Cleaver:2000ds} and $SU(4)\times SU(2)\times U(1)$ models\cite{Cleaver:2002ps}.
 \end{enumerate}

\begin{figure}[!h]
 \begin{minipage}[c]{0.49\textwidth}
  \begin{center}
    \includegraphics[width=\textwidth]{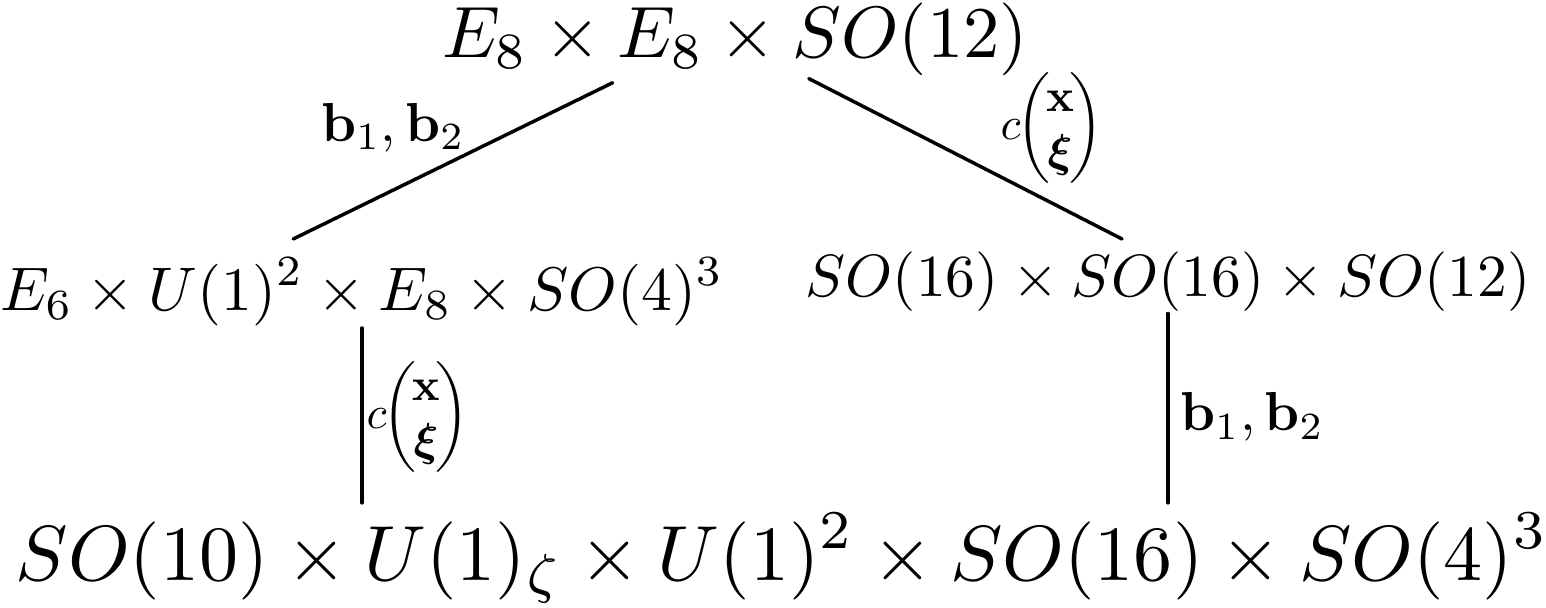}
    \caption{\label{fig:caseibreaking} Symmetry breaking patterns for case (i) models.}
  \end{center}
\end{minipage}
\hspace{0.5cm}
\begin{minipage}[c]{0.49\textwidth}
  \begin{center}
    \includegraphics[width=\textwidth]{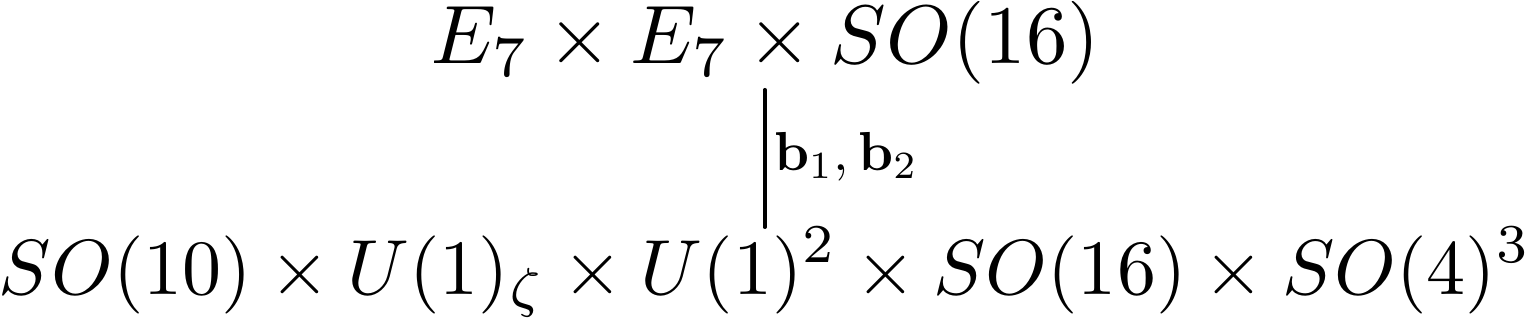}
    \caption{\label{fig:caseiibreaking} Symmetry breaking pattern for case (ii) models.}
  \end{center}
\end{minipage}
\end{figure}

 For the case (i) models, the linear combination of the Cartan $U(1)$s
 \begin{align}\label{eqn:U1zeta}
   J_\zeta=\overline{\eta}^{1\ast}\overline{\eta}^1+\overline{\eta}^{2\ast}\overline{\eta}^2+\overline{\eta}^{3\ast}\overline{\eta}^3
 \end{align}
 is always anomalous \cite{Cleaver:1997rk}.  For the case (ii) models, the $U(1)_{1,2,3}$ individually and, thus, the combination in \eqref{eqn:U1zeta} may be anomaly-free.  This is due to the embedding of the $SO(10)$ GUT symmetry and the $U(1)_\zeta$ combination in \eqref{eqn:U1zeta}: the case (i) models come from an $\mathcal{N}=4$ vacuum with $E_8\times E_8\times SO(12)$ gauge symmetry generated by the basis vectors $\mathcal{B}_{\mathrm{\tiny{I}}}=\left\{ \mathbb{1},\mathbf{S},\mathbf{x},\boldsymbol{\xi} \right\}$\footnote{with $\mathbf{x}=\left\{ \overline{\psi}^{1,\dots,5},\overline{\eta}^{1,2,3} \right\}$ and $\boldsymbol{\xi}=\left\{ \overline{\phi}^{1,\dots,8} \right\}$}, and may follow the breaking patterns in Figure \ref{fig:caseibreaking} as a result of additional basis vectors or changing the GGSO projection operators. In previous literature, the enhancing gauge bosons in the \textbf{x}-sector are always projected out.  
 
The case (ii) models originate from a different $\mathcal{N}=4$ vacuum with an $E_7\times E_7\times SO(16)$ generated by the basis vector set $\mathcal{B}_{\mathrm{\tiny{II}}}=\left\{ \mathbb{1},\mathbf{S},\mathbf{x},2\boldsymbol{\gamma} \right\}$\footnote{with $2\boldsymbol{\gamma}=\left\{\overline{\psi}^{1,\dots,3},\overline{\eta}^{1,2,3},\overline{\phi}^{1,8}  \right\}$} and are broken as shown in Figure \ref{fig:caseiibreaking} upon addition of $\bv{1}$ and $\bv{2}$.  

\begin{figure}[!h]
  \begin{minipage}[c]{0.49\textwidth}
  \begin{center}\includegraphics[width=\textwidth]{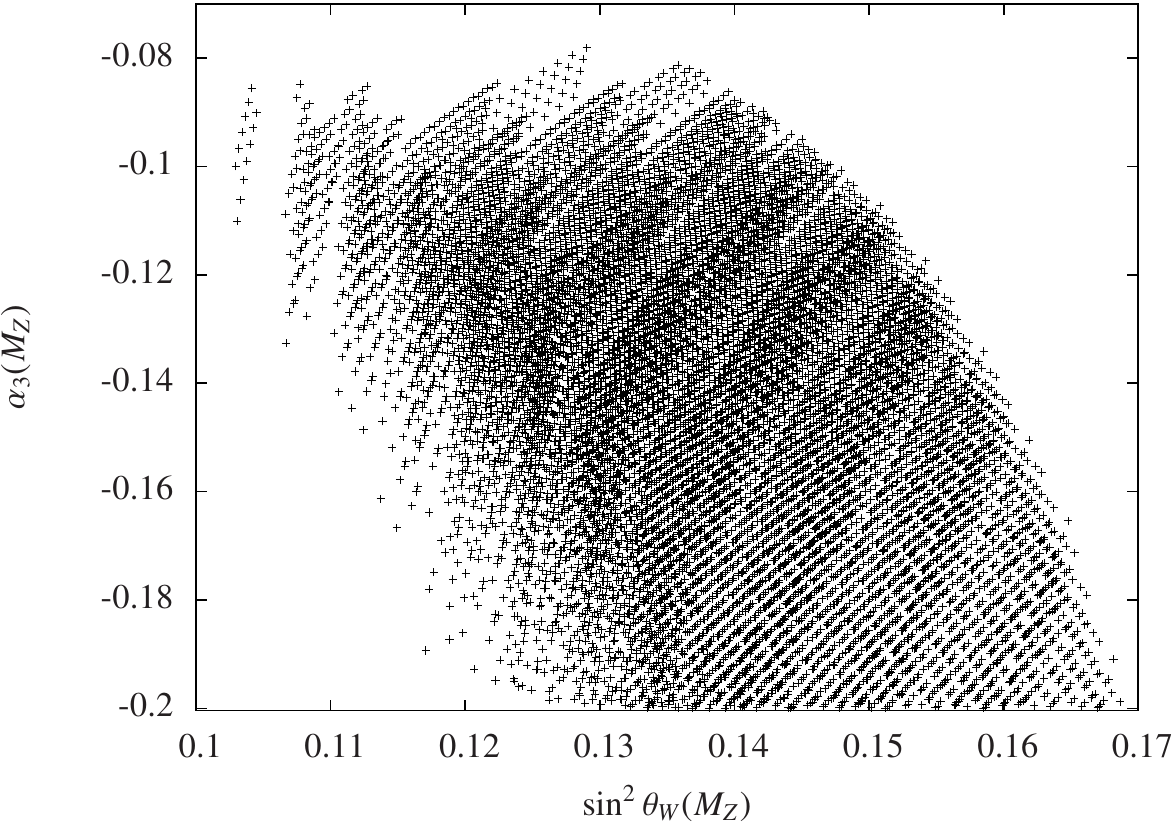}
    \caption{\label{fig:e6gcu} Running the standard model gauge couplings, $\alpha_3(\mu)$ and $\sin^2\theta_W(\mu)$, from the string scale, $M_{\mathrm{\tiny{string}}}$, to $M_Z$, we find that our simple toy model does not match the experimental measurements.  Here we have taken $0<\alpha_{\mathrm{\tiny{string}}}\leq 0.1$ and $2\cdot 10^{16}\leq M_{\mathrm{\tiny{string}}}\leq 5.27\cdot 10^{17}$ GeV\cite{Faraggi:2013nia}.}
  \end{center}
\end{minipage}
  \begin{minipage}[c]{0.49\textwidth}
  \begin{center}\includegraphics[width=\textwidth]{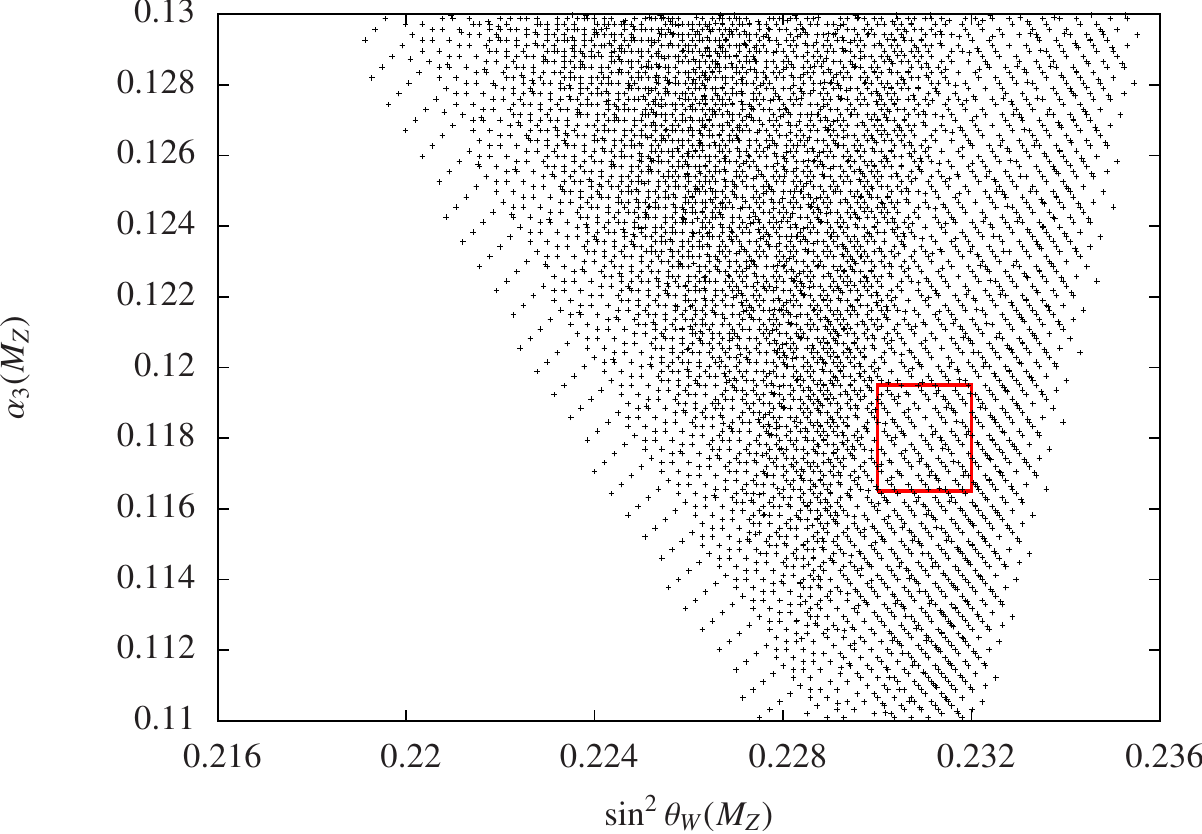}
    \caption{\label{fig:so10gcu} Running the standard model gauge couplings again for the same range of $M_{\mathrm{\tiny{string}}}$ and $\alpha_{\mathrm{\tiny{string}}}$\cite{Faraggi:2013nia}, this time with a string model spectrum containing the full \textbf{27} of $E_6$.  As expected, there are many phenomenologically viable models, as highlighted.}
  \end{center}
\end{minipage}
\end{figure}

The $c(^\mathbf{x}_{\boldsymbol{\xi}} )$ projection removes the gauge bosons that enhance the $SO(16)$ symmetries to $E_8$ and, in addition, projects out either the \textbf{16} or the \textbf{10}$+$\textbf{1} representations in the $\bv{i}$ or $\bv{i}+\mathbf{x}$ sectors, respectively, depending on the $c(^{\bv{i}}_{\mathbf{x}}\!)$ projections.  Thus, the additional states completing the \textbf{27} of $E_6$ are no longer present in the spectrum and so $U(1)_\zeta$ becomes anomalous.  In case (ii) models, there is no $E_6$ embedding.  In fact, the anomalies are avoided due to the components of the \textbf{16} representation having $U(1)_\zeta$ charges of opposite sign.  This was analysed in a toy string-inspired model previously in \cite{Faraggi:2011xu}.  There, we built a toy field theory model, using our the string charge assignments, consisting of the MSSM$+\nu_R$ states, \ie three \textbf{16}s of $SO(10)$.  Additional doublets were required to cancel the $SU(2)^2_{L/R}\times U(1)_\zeta$ anomalies, however, due to these doublets having $Q_\zeta\not\in E_6$, we found no agreement with low-scale gauge coupling data (as shown in Figure \ref{fig:so10gcu}).  Including the full \textbf{27} of $E_6$ in the field theory analysis, and thus having $U(1)_\zeta\subset E_6$, circumvents this problem, due to the cancellation between the additional triplets and doublets in the spectrum when running the renormalisation group equations, shown in Figure \ref{fig:e6gcu}.  

The requirement that $U(1)_\zeta$ is anomaly-free originates in the desire for a low-scale $U(1)$ coming from a string model to act as the proton lifeguard:  the two $U(1)$ combinations orthogonal to $U(1)_\zeta$ are, generically, family non-universal or anomalous; the $U(1)$ combination orthogonal to the weak hypercharge and embedded in $SO(10)$, necessarily breaks at very high scales in order to provide a sufficiently large seesaw scale \cite{Faraggi:1990it}.  Therefore, a light $Z^\prime$ in heterotic string models must be some combination of all of these symmetries and, thus, $U(1)_\zeta$ must be anomaly-free.

\section{A new model}
So, the issue is then not obvious to avoid: the gauge coupling running necessitates an $E_6$ embedding of the $U(1)_\zeta$, however the known string embeddings of such a symmetry require it be broken at the string scale due to anomalies.  A possible solution to this would be to break $E_6$ to a different maximal subgroup, \eg $SU(6)\times SU(2)$, and having the $U(1)_\zeta$ combination embedded in one of the subgroups.  However, to break this to the Standard Model, one requires a Higgs living in the adjoint representation\cite{Bernard:2012vf}; a feature missing from Ka\v{c}-Moody level one worldsheet theories.  

Alternatively, we may construct NAHE-based models that, rather than projecting out the enhancing gauge bosons in the \textbf{x}-sector, they are kept.  This will allow for states in the $\bv{i}+\mathbf{x}$-sectors to remain in the spectrum, thus filling the \textbf{27} of $E_6$.  In order to accommodate this, we add to the NAHE set the basis vectors $\boldsymbol{\alpha},\boldsymbol{\beta},\boldsymbol{\gamma}$, where the visible gauge symmetry is broken by the boundary conditions,
\begin{subequations}\label{eqn:alphabetagamma}\begin{align}
&b\left\{ \overline{\psi}^{1,\cdots5,},\overline{\eta}^{1,2,3} \right\}=\left\{ 1\ 1\, 1\, 0\, 0\, 1\, 1\, 1 \right\} \,\Rightarrow SO(6)\times SO(4)\times U(1)^3,\label{eqn:so64breakingbc}\\
&b\left\{\overline{\psi}^{1,\cdots5,},\overline{\eta}^{1,2,3}\right\}=\left\{1~ 1\, 0\, 1\, 0\, 1\, 1\, 1 \right\}\,\Rightarrow SO(4)\times SO(2)\times SO(2)\times SO(2)\times U(1)^3,\label{eqn:so4111breakingbc}\\
&b\left\{\overline{\psi}^{1,\cdots5,},\overline{\eta}^{1,2,3}\right\}=\left\{{1\over2}{1\over2}{1\over2}{1\over2}{1\over2}{1\over2}{1\over2}{1\over2}\right\}\,\Rightarrow SU(2)\times U(1)\times U(1)\times U(1)\times U(1)\times U(1)^3,\label{eqn:su51breakingbc}
\end{align}
\end{subequations}
with the $U(1)^3$ factor generated by the $\overline{\eta}^{i\ast}\overline{\eta}^{i}$ currents.  The \textbf{x}-sector may then arise in these models either as a separate basis vector or as, for example, $2\boldsymbol{\gamma}$.  The gauge symmetry in \eqref{eqn:alphabetagamma} is generated by the untwisted sector gauge bosons.  This is enhanced by the gauge bosons in the \textbf{x}-sector, to
\begin{align}
  SU(3)\times SU(2)\times U(1)^2\times U(1)_{\zeta^\prime},
\end{align}
where $U(1)_{\zeta^\prime}$ is a rotation of our original combination in \eqref{eqn:U1zeta} and the $U(1)^2$ factor are those embedded in $SO(10)$.  The advantage of such a model is the presence of all the states filling the \textbf{27} of $E_6$ in the spectrum, shown in Table \ref{table:27spectrum}.  The spectrum also contains pairs of heavy Higgs states, 
\begin{align}
  \mathcal{N}+\overline{\mathcal{N}}=\left( \mathbf{1},\mathbf{1},\frac 32, -1, \frac 12 \right)+\left( \mathbf{1},\mathbf{1},-\frac 32, +1,-\frac 12 \right),
\end{align}
required to break this symmetry down to 
\begin{align}
  SU(3)\times SU(2)\times U(1)_Y\times U(1)_{Z^\prime}.
\end{align}

\begin{table}[!h]
\noindent 
{\small
\begin{center}
{\tabulinesep=1.2mm
\begin{tabu}{|l|cc|c|c|c|}
\hline\hline
Field &$\hphantom{\times}SU(3)_C$&$\hphantom{\times}SU(2)_L $
&${U(1)}_{a}$&${U(1)}_{b}$&${U(1)}_{\zeta^\prime}$\\
\hline
$Q_L^i$& $3$ & $2$ & $+\frac 12$ &\hspace{1em}$0$ &\hspace{1em}$\frac 12$\\
$u_L^i$& $\overline{3}$ & $1$ & $-\frac 12$ & $-1$ &\hspace{1em}$\frac 12$\\
$d_L^i$& $\overline{3}$ & $1$ & $-\frac 12$& $+1$ &\hspace{1em}$\frac 12$\\
$e_L^i$& $1$ & $1$ & $+\frac 32$ & $+1$ & \hspace{1em}$\frac 12$\\
$L_L^i$& $1$ & $2$ & $-\frac 32$ &\hspace{1em}$0$ & \hspace{1em}$\frac 12$\\
$N_L^i$& $1$ & $1$ & $+\frac 32$ &$-1$ & \hspace{1em}$\frac 12$\\
\hline
$D^i$& $3$ & $1$ & $-1$ &\hspace{1em}$0$ & $-1$\\
${\bar D}^i$& $\overline{3}$ & $1$ & $+1$ & \hspace{1em}$0$ & $-1$\\
${\bar H}^i$& $1$ & $2$ & \hspace{1em}$0$ & $+1$ & $-1$\\
${H}^i$& $1$ & $2$ & \hspace{1em}$0$ & $-1$ &$-1$\\
\hline
$S^i$& $1$ & $1$ & \hspace{1em}$0$ & \hspace{1em}$0$ & $+2$\\
\hline\hline
\end{tabu}}
\end{center}
}
\caption{\label{table:27spectrum}High scale spectrum and $SU(3)_C\times SU(2)_L\times U(1)_{a}\times U(1)_{b}\times U(1)_{\zeta^\prime}$ quantum numbers, with $i=1,2,3$ for the three light generations. The charges are displayed in the normalisation used in free fermionic heterotic string models \cite{Dienes:1995bx}. }
\end{table}

Due to the survival of the states filling the \textbf{27} representation of $E_6$, it is possible that proton decay mediating operators may be induced. In particular, couplings of the form
\begin{align}
QQD, ud\overline{D}, dND, ueD, QL\overline{D}
\end{align}
are potentially dangerous.  However, the additional states are charged under other symmetries in the string model, some of which are broken at the string scale, and the surviving remnant discrete symmetries may be suitable to suppress these couplings.  Further analysis of this will appear in future work.

\section{Conclusions and outlook}
Here we have presented a novel, string-derived $U(1)$ gauge symmetry that may survive to low scales.  The model presented is free of anomalies and, due to the enhanced spectrum, accommodates the low-scale gauge coupling data.  In addition, the model facilitates the necessary requirements outlined in the introduction and is a viable candidate for a low-scale $Z^\prime$.  

An extension of this work is currently being conducted and an explicit string model, with an accompanying RGE analysis, will appear soon.

\section*{Acknowledgements}
With gratitude to Alon Faraggi and Panos Athanasopoulos for their collaboration during this work.  We thank Nick Mavromatos and King's College, London, for hosting the DISCRETE 2014 Colloquium and their accommodation of an encouraging atmosphere for fruitful discussions.  We're grateful for their kind invitation to present this work, which was supported by the DFG through the grant, TRR33 ``The Dark Universe''.

\end{document}